\documentclass[times]{cpeauth}

\usepackage{moreverb}
\usepackage{comment}
\usepackage[ruled,vlined,linesnumbered]{algorithm2e}
\usepackage{subfigure}
\usepackage{amsmath}
\usepackage{color,soul}
\usepackage[colorlinks,bookmarksopen,bookmarksnumbered,citecolor=blue,urlcolor=blue]{hyperref}

\newcommand{\eg}{{\it e.g., }}
\newcommand{\etal}{{\it et~al., }}
\newcommand{\ie}{{\it i.e., }}
\newcommand{\se}{searchable encryption}

\newcommand\BibTeX{{\rmfamily B\kern-.05em \textsc{i\kern-.025em b}\kern-.08em
\kern-.1667em\lower.7ex\hbox{E}\kern-.125emX}}

\begin{document}

\runningheads{Hoang Pham, Jason Woodworth, Mohsen Amini Salehi}{Survey on Secure Search Over Encrypted Data on the Cloud}

\title{Survey on Secure Search Over Encrypted Data on the Cloud}

\author{Hoang Pham\affil{1}\corrauth, Jason Woodworth\affil{1}, and Mohsen Amini Salehi\affil{1}}

\address{\affilnum{1} High-Performing Cloud Computing (HPCC) Laboratory, School of Computing and Informatics, 
University of Louisiana at Lafayette, LA, USA}

\corraddr{High-Performing Cloud Computing (HPCC) Laboratory\\ School of Computing and Informatics\\ 
University of Louisiana at Lafayette, LA, USA}

\begin{abstract}
Cloud computing has become a potential resource for businesses and individuals to outsource their data to remote but highly accessible servers. However, potentials of the cloud services have not been fully unleashed due to users’ concerns about security and privacy of their data in the cloud. User- side encryption techniques can be employed to mitigate the security concerns. Nonetheless, once the data in encrypted, no processing (e.g., searching) can be performed on the outsourced data. Searchable Encryption (SE) techniques have been widely studied to enable searching on the data while they are encrypted. These techniques enable various types of search on the encrypted data and offer different levels of security. In addition, although these techniques enable different search types and vary in details, they share similarities in their components and architectures. In this paper, we provide a comprehensive survey on different secure search techniques; a high-level architecture for these systems, and an analysis of their performance and security level.
\end{abstract}


\maketitle

\section{Introduction}\label{sec:intro}
As cloud computing becomes prevalent, more cloud-based solutions are developed and widely used in different applications. 
Companies that have adopted cloud storage solutions are reported to gain a competitive edge against those that have not~\cite{marston2011cloud}. 

Cloud computing is favored due to its many advantages, including: convenience and accessibility, consistent back ups to reducing the burden of local storage, and saving capital expenditure on in-house hardware and software maintenance \cite{avram2014advantages}.
Apart from advantages, public cloud storage services are utilized by multi-tenant customers who store large amounts of potentially sensitive data on the cloud.
In addition, using cloud storage implies losing full control over data and delegating it to the cloud administrators. These issues expose the data to potential external and internal attacks~\cite{yan2017flexible, generalsecurity:sohal}, which can be devastating for organizations that rely on confidentiality of their data (e.g financial corporations).  

These problems has made businesses concerned about outsourcing their data to the cloud and making use of its potentials~\cite{nakayama2017effects, generalsecurity:chang}. For instance, a medical center that owns patients' health records cannot outsource its data to a cloud that is vulnerable to attacks, due to legal regulations~\cite{singh2016twenty}. Another instance, is a law enforcement agency that keeps sensitive criminal records and hesitates to use cloud storage. 

One way to overcome the confidentiality problem is to encrypt data on the local premises before outsourcing it to the cloud. While this approach preserves data confidentiality, it hinders data processing. In particular, searching is of paramount importance for outsourced unstructured data~\cite{MEHARWADE2016139}. In fact, when data is encrypted, search systems do not function anymore, because they are unable to compare the query to the encrypted data. 

A na\"{\i}ve approach to enable search on encrypted data would be  downloading all of the data from the cloud, decrypting them, and locally performing plain text search~\cite{Wang2016}. However, with potentially huge data (also called big data) hosted on the cloud and limited network bandwidth~\cite{song2000practical}, this approach remains impractical. Therefore, searchable encryption systems (\eg~\cite{wang2013efficient,moataz2013semantic,sun2014privacy,moh2014efficient}) have been introduced to cope with this problem. These systems ideally allow the encrypted data to be searched without revealing data and the search query. Hence, they relieve concerns about data confidentiality in the cloud.

Efforts to create \se~systems are dated back to early 2000 by Song \etal~\cite{song2000practical}. Since then, numerous research works have been undertaken to enable different types of \se. Although these systems are different in their searching approaches, security level, and performance, they share certain architectural similarities. There are other survey studies over different \se~systems~\cite{Bosch:2014:SPS:2658850.2636328,Poh:2017:SSE:3101309.3064005}. This paper complements those survey studies by providing a comprehensive survey of the existing \se~systems and differentiate them in terms of their searching approaches, security level, and performance. In addition, we provide a generic framework that encompasses the components of \se~systems. As recently more granular search systems are demanded by different industries, we survey domain-specific search systems as well. Finally, we identify the shortcomings of the current research works and recommend avenues for future research and development efforts in the \se~area. In summary, the contributions of this paper are as follows:

\begin{itemize}
  \item Analyzing the commonalities and differences in various types of cloud-based \se~systems.
  \item Providing a generic framework for cloud-based \se~systems. 
  \item Providing survey and categorizing current cloud-based \se~systems.
  \item Surveying domain-specific encryption systems.
  \item Identifying forthcoming challenges in \se~and introducing future research avenues to address them.
\end{itemize}

The remainder of the paper is organized as follow: Section \ref{sec:bg} introduces the background and preliminaries of \se. An overall framework for \se~is given in Section \ref{sec:ovrarch}. Section \ref{sec:security} reviews the security requirements and criteria to assess \se~systems. Section \ref{sec:cat} reviews the categorization of different \se~schemes and Section \ref{sec:shortcoming} evaluates the security shortcomings of the \se~system and also categorizes the existing \se~systems based on their security analysis. Next, Section \ref{sec:app} surveys domain-specific \se~schemes. Finally, Section \ref{sec:conclsn} summarizes the paper and provides avenues for future research in this area. 

\section{Elements of a Cloud-Based Searchable Encryption System}\label{sec:bg}
\begin{figure}[htbp]
\centering
	{\includegraphics[scale=0.45]{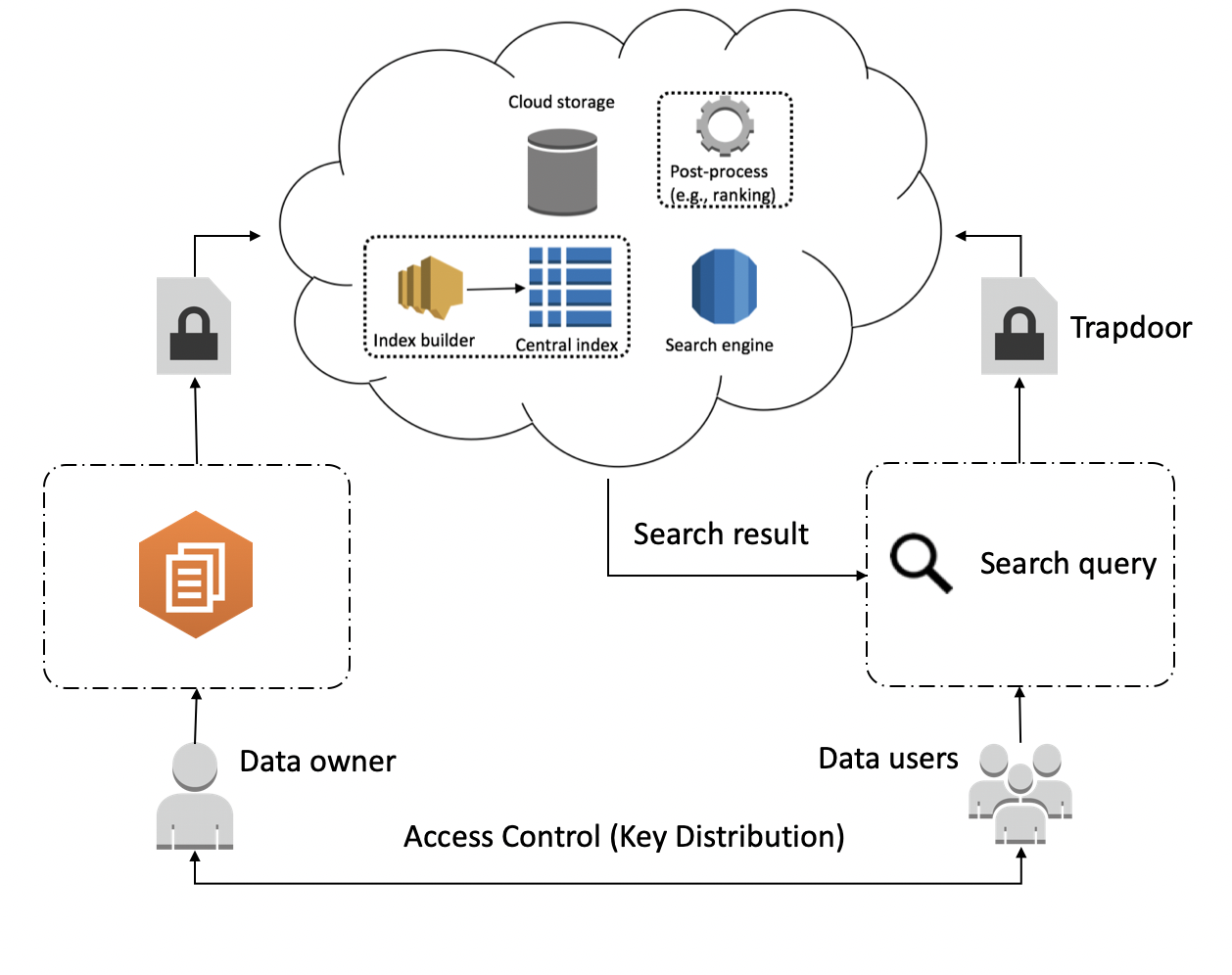}}
    \caption{Main elements involved in cloud-based \se~systems.\label{fig:components}}
\end{figure}

Song \etal~\cite{song2000practical} are one of the pioneers in \se. They provide a system in which a client (\ie data owner) can search over her encrypted data (in the form of emails) on an email server. Once the data owner wants to search some keywords in her emails, she submits an encrypted query (termed \emph{trapdoor}) to the server. The server is in charge of searching over the encrypted data and retrieving the related ones to the owners. 

More recent research works in \se~ (\eg~\cite{sun2014privacy,saleem2017secure,ali2017privacy,amini14,moataz2013semantic,xia2014secure}) describe their systems with similar elements to Song \etal work. In the rest of this section, we introduce these elements in more details.

As depicted in Figure~\ref{fig:components} which is inspired by works in \cite{sun2014privacy,saleem2017secure}, \se~systems are commonly composed of three main elements, as follows: 
\paragraph{Data owner} A data owner sets up the system by granting access control to data users and uploading documents to the cloud.
The data owner posses a collection of $n$ documents $D = \{d_1, d_2, d_3,\ldots, d_n\}$ and wants to outsource them to a remote public cloud server (\eg Amazon and IBM Cloud Storage) for storing or sharing purposes. In order to protect confidentiality of the documents, the owner encrypts the data locally using her authorization key and uploads the encrypted data to the cloud.
\paragraph{Data user} The data users are those who are authorized to search and retrieve the uploaded documents. The data users have an encryption key that is applied on their search queries and creates trapdoors\footnote{For further explanation about trapdoors, please see Section \ref{sec:ovrarch}}. The trapdoors are then sent to the cloud servers to retrieve search results in form of a list of relevant documents (\eg in~\cite{cao2014privacy, fu2015achieving}) or document identifiers (\eg in~\cite{curtmola2011searchable,chang2005privacy}). The data user can be potentially permitted to decrypt the document locally \cite{curtmola2011searchable,chang2005privacy}. It is worth noting that, in practice, a data owner can be a data user too.
\paragraph{Cloud server} The cloud server receives an encrypted collection of documents uploaded by data owners and carries out three main tasks: storing uploaded documents; searching them against trapdoors; and maintaining search data structures updated.

The research works that have been undertaken in cloud-based \se~generally assume cloud servers to be \emph{honest-but-curious}. That is, although the cloud server administrator follows necessary security procedures and does not modify nor delete data files, she is still ``curious'' about the content of the documents. 

\paragraph{} Another component we found common in current \se~systems is trusted computing base. Each \se~system implements this element in its owned way to adapt its purposes and functionality. Although trusted computing base can be combined with other components in the \se~system system, it is worth recognizing and mentioning that in the paper.

\paragraph{Trusted computing base (Gateway)}
Data owner and Data users in a \se~system need data preparation and preprocessing (\eg removing stop words from the search query or extracting keywords from documents \cite{xia2014secure}) in their local domain before proceeding to the Cloud server. The preprocessing constructs another element in the \se~system, termed Trusted computing base (or Gateway) \cite{sun2014privacy,amini14,wang2013efficient}, that includes the client-side application. In particular, job of the gateway is to prepare document in the setup phase for data owner and preprocess user's query in the retrieval phase. The gateway is generally assumed to be trusted and resides in the user's premise. There are two approaches to implement the gateway: \emph{client-end approach} in which the gateway is part of client machine~\cite{li2010fuzzy,fu2017privacy}; and \emph{trusted server approach} in which the gateway resides on separate trusted sever~\cite{sun2014privacy,amini14,wang2013efficient}.

\begin{itemize}
\item \textbf{Client-end approach} The advantage of the client-end approach is that the data is secure from its origin, thus, is safe against connection interception. The drawback of this approach is to impose overhead on the client-side application and potentially affecting its performance. Hence, it is considered inappropriate for circumstances that data users predominantly use thin-client devices (\eg smart phones)~\cite{7841040}.

\item \textbf{Trusted server approach}  The Trusted server approach is generally faster and lends itself better to the edge computing platforms. Although this approach imposes a low overhead on the client machines, it reveals data in the communication channels between the client machine and the trusted server. This approach also includes server provisioning and maintenance costs. In practice, the trusted server approach is appropriate for circumstances where clients' devices fall short in computing power and have limited energy supply \cite{ali2017privacy}. 
\end{itemize}

\section{A General Architecture for Searchable Encryption Systems}\label{sec:ovrarch}
\subsection{Overview}
The architecture of a \se~system has to implement four processes involved with the elements mentioned in the previous section. In this section, we first describe these four processes, then we discuss how they are applied within different elements of the \se~system.

\subsection{Processes Involved in a Searchable Encryption System}\label{subsubsec:keygen}
The processes in a \se~system are namely \emph{Key generation (Keygen), Build-Index, Trapdoor generation,} and \emph{Search}. These processes are enablers of \se~systems and generally have polynomial time complexity~\cite{wang2013efficient,ballard2005achieving}.

\paragraph*{Key Generation (Keygen) Process} This process is in charge of creating key to encrypt plain text documents, and later to decrypt the retrieved documents. The algorithm for this process generates a key based on a set of given security parameters. Probabilistic key generation algorithm~\cite{ballard2005achieving} are commonly used for this process. 

As data owner and data user transmit and store data through the server, there is a need to secure the resided documents in privacy and how to recognized the authorized users to access these data. There are two common methods that encryption documents can be carried out: symmetric and asymmetric encryption.

\begin{itemize}
\item Symmetric Encryption: In this methods, both data owner and data user share the same secret key. This secret can be used to both encrypt and decrypt the document. In other words, this key is to create unclarity noise in the documents to the unauthorized user while the authorized data user can use that shared key (or a computable \"inversed\" key to defuse and remove the uncertainty in the document \cite{fu2016enabling}
\item Asymmetric Encryption: Known as Public Key Encryption. This cryptography includes two different key: public key and private key which are used to encrypt and decrypt the document \cite{boneh2004public,boneh2007public}. More specifically, the data owner would use one of his keys for encryption the document and the other to decrypt. Since the two key are completely different and there is no computationally correlation between them, even if the encryption key (public key) is compromised, the attacker can not get the message content without the private key.
\end{itemize}

\paragraph*{Build-Index} 
Searchable encryption systems commonly utilize an index structure to keep track of occurrences of keywords in documents. The process of initializing this index, called \emph{Build-Index}, takes key $K$ from \emph{Keygen process} the and a collection of documents $D$ as inputs. Then, it extracts keywords from the documents and insert them into the index structure. 

This \emph{Build-Index} process is used by the data owner to generate a secure and searchable strucure that enables search over the encrypted data. An index structure is generally implemented in form of a hash table~\cite{cash2014dynamic}, meta data (markup)~\cite{saleem2017secure}, or inverted index~\cite{7841040,woodworth2018s3bd} where each unique keyword is mapped to document identifiers it appears in.

\paragraph*{Trapdoor Generation} This process is used by data users to form search queries. It encrypts the user's search query using a key that is compatible with the Build-Index key $K$. 

It is possible that the search query preprocessed (\eg expanded) by the Trapdoor Generation process~\cite{fu2017privacy,xia2014secure,moataz2013semantic}. Then, the encrypted Trapdoor is sent  to the cloud server.

\paragraph*{Search} After receiving the trapdoor, the server runs the search procedure to match documents that contain the set of keywords in the trapdoor. Next, the results are sent back to the client. 

\subsection{General Architecture for Searchable Encryption Systems}
\begin{figure*}[htbp]
\centering
	{\includegraphics[scale=0.43]{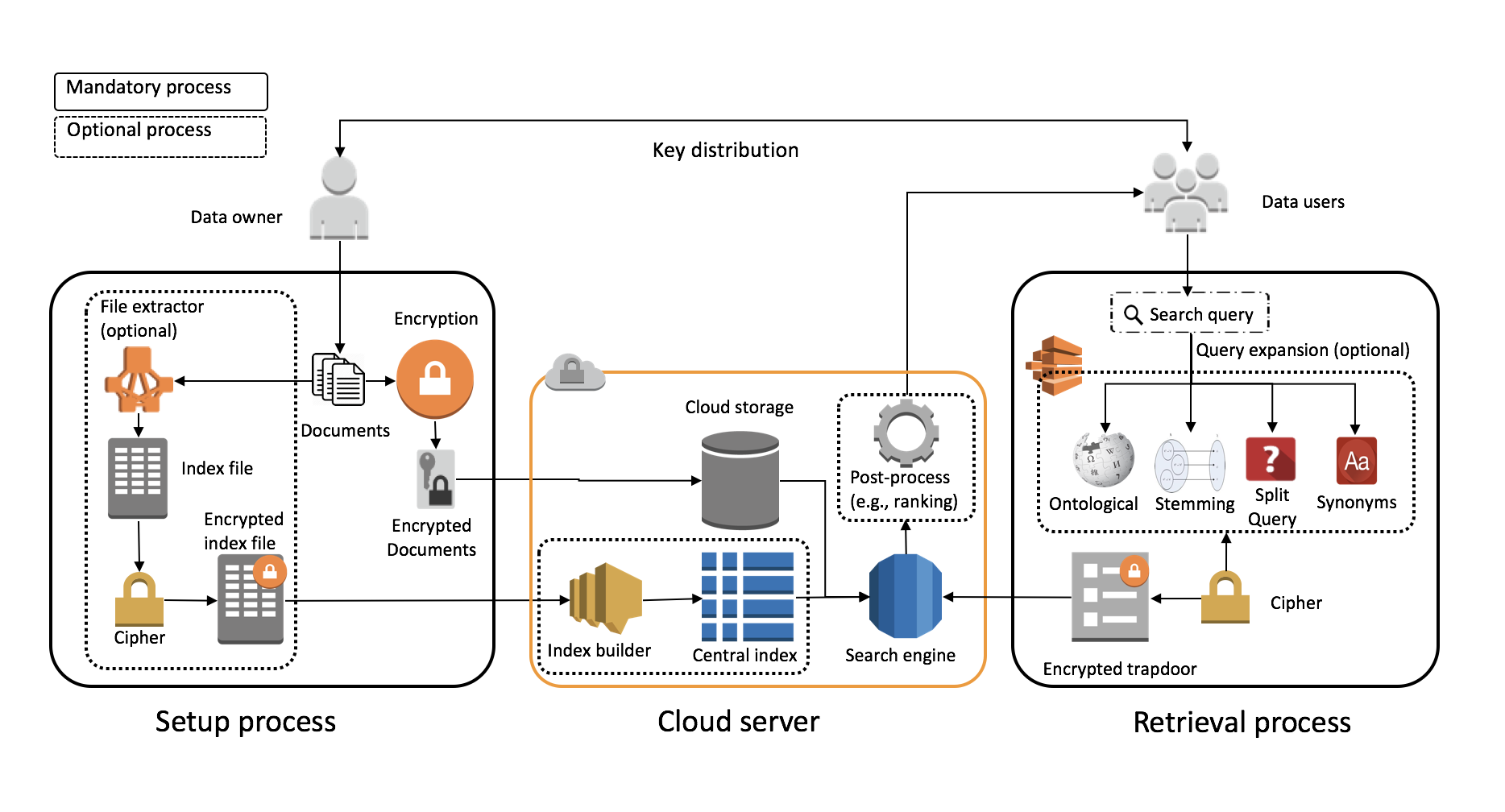}}
    \caption{Architectural Overview of Cloud-based Searchable Encryption Systems.\label{fig:ovrarch}}
\end{figure*}

The general architecture for a cloud-based \se~system is depicted in Figure \ref{fig:ovrarch}.

The system architecture consists of two main mechanisms, namely \emph{Setup} and \emph{Retrieval}. 
The main job of the Setup mechanism is to prepare documents for searching by data users. Upon receiving a search query from the data user, the job of Retrieval mechanism, is to perform search on the dataset, find matching documents and send the results back to the data user. 

In the next subsections, we elaborate on how these mechanisms operate.
\subsubsection{Setup Mechanism}
Before sending documents to the cloud, in the Setup mechanism, first, the useful information from the documents are extracted. The type of extracted data depends on the type of search system (\eg keyword search~\cite{wang2012enabling,wang2010secure} versus semantic search~\cite{sun2014privacy,moh2014efficient}). 
Then, the extracted data and documents are encrypted and sent to the cloud server. 

The data owner initiates the search system through the \emph{Keygen} process (see Section~\ref{subsubsec:keygen} for more details). The generated key is necessary to encrypt documents before outsourcing, and decrypt downloaded documents. 

In some \se~systems, the Setup mechanism also includes creating an ``Index structure'' \cite{sun2014privacy,moh2014efficient,xia2014secure,wang2012enabling}. The Index structure is also known as ``meta data'' \cite{saleem2017secure} or ``identify keywords'' \cite{boneh2004public,goh2003secure}). The index comprised of keywords that represents the essence of each uploaded document. Alternatively, some other \se~systems do not rely on the index structure. They directly encrypt each keyword individually and form an encrypted searchable document \cite{song2000practical}. 

The \se~system then encrypts the documents' contents as well as the index structure (if it exists), before sending them to the cloud server. 

In systems that data owner and data users are separated entities, the data owner needs to distribute the public-key to the data users. The keys are utilized by data users to create trapdoors that are compatible with the encrypted uploaded data. Methods such as public key cryptography~\cite{cash2014dynamic} or broadcast encryption \cite{wang2010secure} are commonly used for key distribution.

\subsubsection{Retrieval Mechanism}

After setup, the system is expected to have a collection of files ready to be searched over. Data users or owners can submit a search query, defined as a set of keywords $W=\{w_1, w_2, \ldots, w_n\}$. 

Trapdoor is produced using $W$ and the keys data user owns.
Some systems (\eg~\cite{cao2014privacy,xia2016secure}) also apply pre-processing of the search query in producing Trapdoor. Once Trapdoor is produced, it is sent to the cloud server.

The cloud server includes a search engine that carries out the search process. In the systems that rely on an index structure, the index is used to match Trapdoor against index entries to find relevant documents. At the end, the list of results, which includes matching documents or their identifiers, are sent back to the user. Upon receiving the result list, the data user can request to receive (\ie download) and decrypt the documents, if they are authorized. 

It is worth mentioning that during the Retrieval mechanism, the cloud server could learn minimal information about the documents \cite{wang2010secure, xia2014secure}. In the next section, we provide further details on security aspects of cloud-based \se~systems.

\section{Security Requirements and Shortcomings of Searchable Encryption Systems}\label{sec:security}
\subsection{Security Criteria of \se~Systems}\label{subsec:crit}
It is crucial for \se~systems to prove that they can preserve confidentiality of the user's data and prevent information leakage. Therefore, the resistance of \se~systems should be verified against possible internal or external attacks on an untrusted server. More specifically, the server should not be able to learn anything about the original data, from the cipher text or the search process. Song at el \cite{song2000practical} defined 3 security properties that every \se~system should maintain:

\begin{enumerate}
\item \emph{Controlled searching}: Unauthorized users should not be able to search in the server. In \cite{goh2003secure, ding2012efficient}, an unauthorized user cannot submit a query to the server unless she has the secret key to generate a trapdoor. On the server side, the data must be kept in encrypted manner. Also, processing of the data must be done without decrypting the data. Thus, without a trapdoor, nothing can be returned in response to an unauthorized search request.

\item \emph{Hidden query}: this technique hides the unencrypted query from the untrusted server. Every \se~system should be able to mask or encrypt the content of the search query to avoid the possibility of the server inferring the content from the search results. The untrusted server can only learn about the relationship of a secure query to a set of document identifiers, not what that set of documents are about. 

Without an encrypted query, an attacker can submit numerous queries to the \se~system. Then, by analyzing the search results, documents' contents can be inferred \cite{goh2003secure}. In \cite{goh2003secure}, the authors present secure index (\eg Z-idx) that  does not reveal the actual search query, by hashing it into an irreversible trapdoor, before sending it to the server. Hence, the server cannot learn any information from the query. Similarly, in \cite{homomorphic:indexing:ren}, Ren \etal~implement a secondary homomorphic encryption on top of deterministically encrypted query terms to further obfuscate the original query.

\item \emph{Query isolation}: In the search process, the server should know nothing, except the search results \cite{song2000practical}. In \se~systems (\eg \cite{ding2012efficient}), if there is a match between the query and the index, the server can locate the related documents and return them to the data user. However, since the data on the server is encrypted and secret key is not stored in the server, the server cannot understand anything other than the search result.
\end{enumerate}

\subsection{Shortcomings of Searchable Encryption Systems} \label{sec-shortcoming}
The security requirement means that the system needs to adopt methods to protect the data. For example, documents and auxiliary indices are encrypted using the secret key so that only a user with the valid key can learn about the content, thus, its privacy and confidentiality from the untrusted server are protected. Even with cipher text, there are prevention measures against different hacking techniques, such as statistical attacks~\cite{kocher1998introduction} and keyword-distribution attacks~\cite{kocher1998introduction}, that aims at deriving the correlation of the cipher text and the plain text. These prevention measures contribute to the complexity of the searching mechanism.

The complexity of the prevention measures compels the \se~system to trade-off between performance and security. In \se~systems, the \emph{access pattern} and \emph{search pattern} are the most common security factors to consider for these trade-offs. 

\paragraph*{Search pattern} is defined as any information or pattern that an attacker can use to derive, if random queries are related to a keywords~\cite{curtmola2011searchable}. In some of the earlier research works (\eg~\cite{song2000practical,chang2005privacy,goh2003secure}), the \se~system reveals the search pattern to in favor of efficiency in the search operation. In a \se~system by Curtmola \etal~\cite{curtmola2011searchable}, the trapdoor is created using deterministic encryption, hence, the system leaks the search pattern. Similarly, research works undertaken in \cite{song2000practical,chang2005privacy,goh2003secure} compromise search pattern to improve search efficiency.

\paragraph*{Access pattern} is defined as any information that the attacker can use to determine the frequency at which files are accessed or associate to the query. For example, an observant and patient attacker could sniff the network connections to understand which files are searched for most frequently, thus, determining the most important documents in a dataset. 

In the context of \se, Goldreich \etal \cite{goldreich1996software} achieved fully secure \se~using Oblivious RAM (ORAM) to hide the access pattern. ORAM periodically \emph{shuffles} the data blocks so that the user avoids accessing the same data (memory) block for the same retrieved data. However, in this work, the initial setup phase is computationally expensive and the search phase requires logarithmic rounds of interaction between the user and the server to narrow down the search results. This makes it a burden on the bandwidth and not suitable for large scale datasets.

Another work aimed at hiding the access pattern has been carried out by Boneh \etal \cite{boneh2007public} using Private Information Retrieval (PIR) technique. In a system consists of $n$ replicated servers, the user sends homomorphic encrypted queries~\cite{banawan2018capacity,boneh2007public} to the $n$ servers to retrieve data blocks from the servers that collectively make up the result \cite{mayberry2014efficient}. The system has to \emph{touch} all of the replicated servers, thus, imposes a significant communication overhead to retrieve one single query. 

To provide an efficient \se~system, Song \etal \cite{song2000practical} resorted to the weakened security guarantee, \ie revealing the access pattern and search pattern but nothing else. 
Later works (\eg\cite{curtmola2011searchable,chang2005privacy, homomorphic:indexing:ren}) attempt to find a proper method to balance a reasonable access and search pattern leakage and improve the search performance.

We should note that even with these security compromises, it is still difficult for an attacker to derive helpful information from the dataset. In fact, the attackers cannot learn anything beyond the importance of certain documents.

\subsection{One-to-Many Order Preserving Symmetric Encryption (OM-OPSE)} \label{omopse}
In \se~systems, data users generally need to see the search results ordered by their relevance to the search queries. As such, a \se~system needs a method to assign relevance score to each found document and rank them for the user. 

One way to achieve ranking of the search results is to assign an importance value (\ie weight) to the  extracted keywords in the index structure. The weight of an extracted keyword can be obtained based on its importance in the document \cite{fu2016enabling,fu2017enabling}. These weights are typically hidden using deterministic encryption methods \cite{li2015security}. However, because deterministic encryption is a one-to-one mapping of plain text to cipher text, this encryption method can potentially leak some information about the documents \cite{li2015security}. In fact, an internal attacker can potentially deduce the distribution of encrypted scores in the data collection.

Given the context background of the corpus, the attacker can utilize the retrieved score distribution to gain knowledge about the documents or even break the encryption of the system \cite{wang2012enabling}. To prevent this, certain research works have modified the Order Preserving Symmetric Encryption (OPSE) which is a deterministic encryption method that preserves numerical order of the plain text \cite{boldyreva2009order}.

Wang \etal \cite{wang2010secure} and Sun \etal \cite{sun2014privacy} utilized OPSE to create a more efficient approach (known as one-to-many order-preserving mapping) to increase obfuscation of the original encrypted scores while maintaining plain text order. Instead of mapping the plaintext score to a single encrypted document, the encrypted score is assigned within a randomly appointed bucket within a predefined range. Therefore, the randomness of score distribution increases and the probability of being predicted decreases. Several research works have demonstrated that modifying OPSE is ``as-strong-as-possible'' encryption technique \cite{wang2010secure,sun2014privacy}. 

\section{Taxonomy of Searchable Encryption Systems}\label{sec:cat}
\begin{figure}[h!] 
\centering
\includegraphics[scale=0.4]{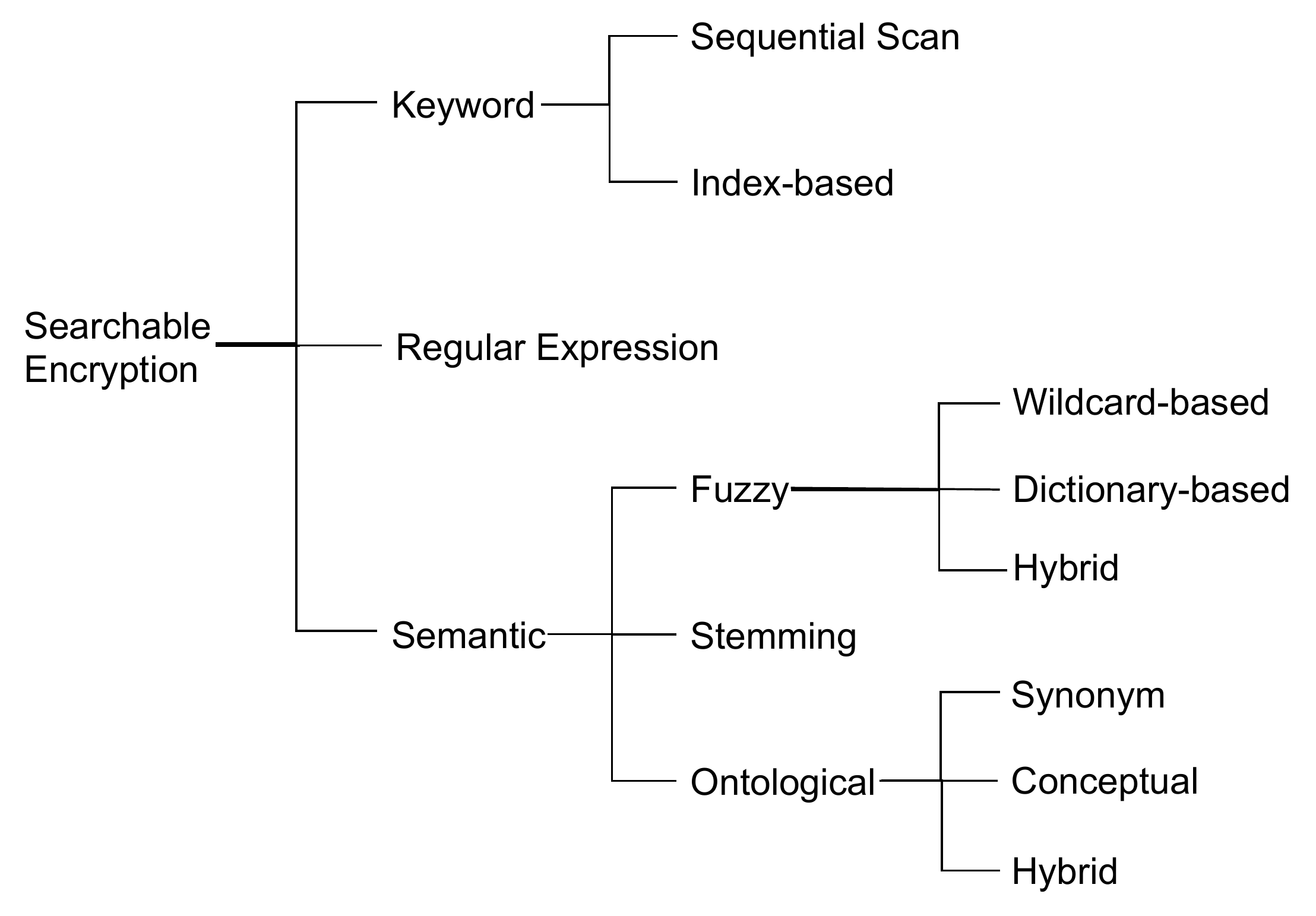}
	\caption{Taxonomy of the current \se~systems.\label{fig:tax}}
\end{figure}

There have been various research efforts to enable different forms of search operation on encrypted data. However, a lot of them are rooted from some general approaches. The goal of this section is to provide a comprehensive taxonomy of the current \se~systems and supply an overview of the conducted research in this domain. The taxonomy of \se~systems is provided in Figure \ref{fig:tax}. As we can see in this figure, the \se~systems can be broadly categorized based on the type of search they perform into three main types, namely Keyword Search; Regular Expression Search; and Semantic search.

\subsection{Keyword Search}
\subsubsection{Sequential Scan Keyword Search} 
Song \etal \cite{song2000practical} pioneered the idea of user-side encryption and keyword-based search over the encrypted documents. The user intends to retrieve encrypted documents that contain her search keyword. In the proposed scheme, each keyword $W_i$ of a document is encrypted independently in two encryption layers. Firstly, they pre-encrypt the keyword $W_i$ with $E(W_i)$ into $n$ bits. 
This is then split into two parts: the right part ($R_i$) consists of $m$ bits and left part ($L_i$) consists of $n-m$ bits. 

Secondly, the left part is encrypted with a stream cipher which has characteristics to check for matching using XOR method.

When a user requests documents that include a set of interested keywords, she submits the encryption version of each keyword of the query: $L_i$ and $k_i$, the server performs XORing matching on $n-m$ bits of each cipher text to see whether $C_i$ XOR $L_i$ is of the form ($S,F_ki(S)$) for some S \cite{song2000practical}. In this system, since every word is independently encrypted, to find all the matches, the server has to follow the above steps word-by-word for the entire document. For this reason, this method of searching is known as sequential scan. 

Another work in Sequential Scan category is PEKS (Public key Encryption with Keyword Search) \cite{boneh2007public}. PEKS encrypts each keyword in the uploaded document based on a public key using Bilinear Diffie-Hellman~\cite{boneh2007public} or Trapdoor Permutation~\cite{boneh2007public} method. 
In the client side, the data owner uses her private key to verify if her query keyword occurs in a document. In this method, the server needs to scan every cipher text (\ie keyword) of each document to find the occurrences. The method is computationally expensive because the search cost is proportionate to the number of the documents in the dataset. 

There is not many dominant research works that also follow this fashion due to its limitation of time complexity to search through the whole dataset size, given the fact of growing volume of data in current storage. More advanced and efficient \se~systems are mentioned later.

\subsubsection{Index-Based Keyword Search}  
In order to deal with the inefficiency of the Sequential Scan keyword search, Song \etal \cite{song2000practical} proposes a data structure called an ``index'' that contains a list of keywords mapped to its original document using a document pointer (also known as document identifier \cite{li2010fuzzy}). The keyword $W_i$ is encrypted as $E(W_i)$ and the document identifier is encrypted using a key resulting from a key generator function $f_k$ with input of the hashed keyword $E(W_i)$. 

In the Index-based keyword search systems, instead of searching sequentially (word-by-word) for every document, the system only needs to check the index structure for the interested keywords. From the results, the system retrieves the document identifiers and sends them back to the client. 

Goh \etal \cite{goh2003secure} was one of the first works to introduce the secure index-based search method. In this work, the system creates an index for each document before encrypting and uploading it. The index consists of keywords that are considered relevant to the document.  
There are two main ways to construct the secure index: IND-CKA~\cite{goh2003secure} and the efficient variant construction called Z-IDX~\cite{goh2003secure}. 
Both of these methods use Bloom Filter~\cite{goh2003secure} as an index for each document to track its keywords. At the search time, the system creates search trapdoor and uses the Bloom Filter to check if the trapdoor is contained in the dataset.

Ren \etal \cite{homomorphic:indexing:ren} directly extended the two-layer encryption idea originally presented by Song \etal into a hashed index representation, seeing significant speed increases. In addition, they introduced another layer of randomized XOR-homomorphic encryption on the search query to further obfuscate the query to adversaries watching the network stream.
Liu \etal \cite{liu2015efficient} presented a \se~method with multiple data sources where the cloud index is composed of multiple Index(es) from different sources. Research works such as \cite{wang2012enabling,wang2010secure} introduced a similarity score for keyword search using formulas based on the term frequency of the keywords. Introducing similarity score allows the result list to be ranked for the end users. We will review their scoring and ranking keyword against document and query in details in Section~\ref{F-sec-cat:ontological}.

Earlier works from \cite{goh2003secure,song2000practical,boneh2007public,curtmola2011searchable} using index-based keyword search with their owned methods to perform single keyword or disjunctive keywords search on encrypted documents. However, as the requirements of data user are growing and in need for more accurate search result, different techniques are introduced on to the index to support multi-keyword and conjunctive keywords search. MRSE \cite{5935306} is one of the first works propose solution to that demand. Cao \etal \cite{5935306} define a dictionary consisting of all the keywords and each keyword has a defined location in the dictionary. Data file and search query are represented by binary vectors which, then are used in "inner product computation" to measure the similarity of the data file and the query. Cao \etal \cite{5935306} also apply their owned internal ranking to return relevant ordered result. Later on, Z. Xu \etal \cite{6413690} improve MRSE in so its fixed dictionary can be extended dynamically. Their works also improve the ranking algorithms by using access frequency in weighing the matched data file.

\subsection{Regular Expression Search}
One expansion to the keyword-based \se~is to allow users performing regular expression search on encrypted data. A preliminary approach by Song \etal \cite{song2000practical} proposes to create all possible variations of a given regular expression. For instance, for $ab[a-z]$ query, it generates all 26 possible search queries that are ${aba, abb, \ldots, abz}$. This approach only works for simple regular expressions with and is not scalable for those with high degree of variability, \eg $a^*b^*$.

RESeED \cite{amini14,aminispe17} is a regular expression search system for encrypted data. RESeED operates based on two data structures: a \emph{Column Store}, which is an unencrypted inverted index, representing keywords and the documents they appeared in; an \emph{Order Store} which is a fuzzy (hashed) representation of keywords within a document. For a given search phrase, RESeED builds a Nondeterministic Finite Automaton (NFA)~\cite{amini14}. The NFA is then partitioned into sub-NFAs that can be matched against keywords in the Column Store. For the documents found in the previous step, their Order Store is checked to confirm the keywords are in the same order of the regular expression. 

\subsection{Semantic Search}
Searchable encryption solutions providing keyword or regular expression-based search abilities are useful when users exactly know what keywords they are searching for in the documents.
However, with a growing collection of documents and the emergence of big data, the data users may not remember exact keywords they want to retrieve, or they might want to search for documents that are more broadly related to a topic \cite{7841040,woodworth2018s3bd}. 

For instance, in a hospital with encrypted medical records, a doctor may desire to search for records using the query ``heart disease''. While the doctor is interested in documents containing the exact query terms, she is also interested in documents with semantically related terms (\eg ``heart attack'' or ``chest pain'').  Therefore, semantic search ability is needed to return documents related to terms in the query and to avoid redundant searching attempts.

Many research works have been undertaken to enable different forms of semantic \se. As we can see in the taxonomy (Figure~\ref{fig:tax}), these semantic search systems can be further categorized into three main types, namely Fuzzy Keyword Search; Stemming Search; and Ontological Search. 

\subsubsection{Fuzzy Keyword Search} 
A Fuzzy Keyword Search improves the usability of a system by searching for close matches to the query if it fails to find sufficient matches for the exact query. Although these systems are categorized as search systems, they may not be directly used for search purposes. Fuzzy search can be particularly used to make a system tolerant to user's typos \cite{fu2016toward}.

Fuzzy keyword search systems work based on \emph{edit distance} that measures the similarity between two strings $s_1$ and $s_2$ \cite{wang2014privacy}. It is defined as the number of string operations needed to transform $s_1$ to $s_2$. The possible string operations are insertion (insert a character into a string), substitution (replace one character with another in a string), and deletion (remove a character from a string) \cite{li2010fuzzy}. 
Let $D = \{d_1, d_2,\ldots, d_n\}$, a collection of documents stored on an untrusted third-party server (\eg cloud); $W = \{w_1, w_2, \ldots, w_n\}$ a unique set of keywords with a fixed edit distance $d$; and $(s,k)$ a search trapdoor with threshold $k \leq d$. Then, a fuzzy keyword search system outputs a list of documents that possibly contain keyword $w$, if $w \in W$; else return document where $ed(w,w_i) < k$.

Li \etal \cite{li2010fuzzy} provide a fuzzy keyword search system that injects all possible words $w_i$ that satisfy $ed(w,w_i) < d$ where $w$ denotes the extracted keyword. For example, the fuzzy list of the keyword CAR is \emph{\{ACAR, CAAR, CARA, $\ldots$, CARZ\}}. This list of fuzzy variants is sent to the server where the index structure is located and includes those keywords and their associated documents. Similarly, in the Retrieval phase, the query is augmented with fuzzy variants, sent to the server to be matched, and a list of documents is returned. However, this approach is computationally expensive. For instance, if $d = 3$ the number of possible variants is $\frac{4}{3}\cdotp k^3 \cdotp 26^3$. To improve this, the authors \cite{li2010fuzzy} propose a wild-card-based technique that inserts wild card (*) to represent the fuzzy character or omit a letter from $w_i$. For instance, the fuzzy key set of CAR is $\{*CAR, C*AR, CA*R, CAR*, CAR\}$ and 
$\{CAR, AR, CR, CA\}$. These techniques significantly reduce the size of the index structure, thus, improve the search time.

In a later study, Liu \etal \cite{liu2011fuzzy} proposes a Dictionary-based Fuzzy search method. The method takes a dictionary and a predefined edit distance value and generates a set of fuzzy keywords for each keyword in the search queries. However, instead of inserting a wild-card, this technique only injects words that exist in the given dictionary. At the search time, the query goes through the same process to get the fuzzy set extension from the dictionary and encrypts it before sending it to the server to be searched. 

In fuzzy keyword search, the system needs to search for the entire list of fuzzy keywords which imposes an extensive overhead for a large edit distance. Therefore, to further enhance the fuzzy searching performance, Wang \etal \cite{wang2013efficient} build a tree from the fuzzy keyword set that reduces the search to $O(log(n))$ of the fuzzy list's size. 

\subsubsection{Stemming Search}
Utilizing the Fuzzy Keyword Search makes \se~systems more resistant to minor typos, but in many cases does not exactly cover the semantic perspective. In fact, two keywords having a close lexicographical distance does not necessitate that they are semantically related. Different words (e.g. ``student'' and ``studying''), despite having a large edit distance (4), are highly semantically related. The stemming search method aims at solving this problem based on the belief that semantically related words tend to start from the same root (stem). 

The model is the same as other \se~systems: a user encrypts documents and extracts keywords as an encrypted index. The difference in this system is the additional step it takes to convert the keyword set into a set of stem words. When a user searches for a query, the query keywords are replaced by its stem words \cite{moataz2013semantic}. There are three ways to extract the stem of a word:
\begin{itemize}
\item Affix stripping: This method applies well-known stemming algorithms to find the stem of a word. Prominent methods include the J.B. Lovins \cite{lovins1968development} and Porter stemming algorithms \cite{porter1980algorithm}. These algorithms remove the suffix and prefix to get the root of a word. However, these algorithms require the language knowledge and can be computationally costly.

\item Statistical Stemming: known as n-gram stemming algorithm to statistically find the frequency of contiguous sequence of $n$ items from a given sequence of text in the whole document \cite{wiki:xxx}. The lowest frequent n-gram is considered to be the stem (also called root). 

\item Hybrid: Utilizing both of the aforementioned methods to find the stem. 
\end{itemize}

Both the uploaded documents and the search query extension go through the same process to get the stemming of the extracted keywords. These stemming keywords are stored in the index on of the third-party server (\eg public cloud) and used for in the search process.

\subsubsection{Ontological Semantic Search} \label{F-sec-cat:ontological}
Fuzzy keyword and stemming \se~methods cannot truly capture the semantic essence in searching. For example, if a user intends to search for “robbery”, she is interested to see results about “burglary” or “break in” as well. However, neither the fuzzy presentation nor the stemming method can capture this type of semantic. In fact, the semantically related words neither share the same stem nor have a close edit distance. To resolve this problem, ontological semantic search was invented to find more meaningful and related data to the original query \cite{7841040,woodworth2018s3bd}. As we can see in the taxonomy of Figure~\ref{fig:tax}, Ontological semantic search can be achieved using synonym semantics, conceptual semantics, or a combination of the semantics. In this part, we explore these schemes, however, we first to revisit some of the preliminary concepts that are enablers of Ontological Searching:
\begin{itemize}
\item \emph{Semantic Relationship:} Psycholinguists, Church \etal \cite{church1990word} propose that the word association can be inferred from statistical description of semantic relationship between words through the co-occurrence of words. To calculate the similarity score between two keywords, Sun \etal \cite{sun2014privacy} use data mining methods to effectively find out the co-occurrence degree between terms in a dataset. For two string $x$ and $y$, the similarity score information $I(x,y)$ is defined based on Equation~\ref{eq:sim}.
\begin{equation}\label{eq:sim}
I(x,y) \equiv \log_2(\frac{P(x,y)}{p(x)p(y)})
\end{equation}

where $P(x,y)$ is the probability that $x$ and $y$ appear together; and $p(x) and p(y)$ are the probabilities that $x$ and $y$ appear independently in the collection. Higher values of the similarity score express the more relevance between $x$ and $y$.

Another method to measure the similarity is based on Cosine similarity \cite{sun2014privacy,6674958} that can be calculated using the vector presentation of the documents and the queries to get a relevant score of the closeness between the queries and documents. 

\item \emph{Inverted Index:} It is a data structure that maps each unique keyword to the documents that contain them. This structure keeps track a list of keywords throughout the whole dataset, each keyword is associated with list of documents that it appears in. For some work \cite{sun2014privacy,saleem2017secure,wang2012enabling} in order to further assist ranking functionality, a normalized numerical relevance score often (value between [0,1]) is also given along with each document to indicate the relevance of it to the key . 

\item \emph{Ranking function:}
In a large dataset, commonly, an abundant number of documents match a certain semantic search query with different degrees of relevance. Therefore, it is necessary for the user to receive the list of documents based on the relevance order. This introduces the need for a ranking function that measures the relevance of matching document to the search query. The most common ranking function is known as $TF \times IDF$ (term frequency, inverted document frequency)~\cite{sun2014privacy,moh2014efficient} where $TF$ indicates the significance of that keyword in the document and $IDF$ indicates the significance of the keyword over all documents in the dataset. 
\end{itemize}

Different methods have been developed to measure the relevance of a given keyword in \se~systems. Woodworth \etal \cite{7841040} extend Okapi BM25 standardized text retrieval method~\cite{7841040} which use term frequency and inverse document frequency to calculate the semantic relevance of a given query to a document. These score later on is used to rank documents in the result set. 

Sun \etal and Xia \etal \cite{sun2014privacy,xia2014secure} propose different mechanisms for semantic \se~by extending the query keywords and rank the result set. The data owner constructs \emph{metadata} for each document and sends the encrypted \emph{metadata} to a trusted server (\eg a private cloud) to build an inverted index and a semantic relationship library (SRL). The inverted index is then sent and reside in the public cloud. Upon receiving a search query, the trusted server extends the query using the semantic relationship library to get ontologically related keywords and synonyms. Then, it encrypts the extended keywords and sends them to the public cloud to look up the index structure. The returned documents are ranked based on the relevance ranking functions and sent back to the user. 

In another study, Moh \etal \cite{moh2014efficient} introduce three schemes to learn the semantic meaning of a search query in order to produce accurately related result. Particularly, they propose Synonym-Based Keyword Search (SBKS), Wikipedia-Based Keyword Search (WBKS), and a combination of the two schemes called WBSKS. In \emph{SBKS} scheme, besides extracting the important keyword of the document, its synonyms that represent the semantic of a document are collected. The encrypted version of these keywords and synonyms are sent to the cloud to form a searchable index.
Similarly, the search query is also extended with its synonyms to form the trapdoor. Then trapdoor is compared with the cloud index structure to find documents that are semantically matching the search query. In \emph{WBKS} scheme, a pre-defined set of Wikipedia articles (WKS) are collected and a vector representation (VR)~\cite{moh2014efficient} of them are created using the term frequency and inverse document frequency ($TF \times IDF$) technique. 

When document is uploaded, its VR is then compared against the VR in WKS and the obtained score is stored in the index structure. In the searching phase, the user query is converted to a VR and is compared to the WKS library using cosine similarity method. This score is then added to the trapdoor to let the server know how semantically related the query is to the WKS library. The cloud server computes the cosine similarity score of the trapdoor and the existing entries in the index to create a ranked list of documents and returns to the user. In \emph{WBSKS} scheme, both SBKS and WSKS techniques are used. SBKS is used to expand the query in the Search phase and WSKS is used to construct the expanded index of the uploaded documents of the Setup phase.

Another way to get the semantic meaning is by using WordNet \cite{saleem2017secure}. WordNet is a tool created by Princeton University that contains a dictionary including word definitions and their synonyms. For example, Yang et. al \cite{semsearch:CCPE} utilize WordNet to construct a semantic keyword set. Before the query is encrypted and searched with, each query keyword is expanded with the provided synonyms.

Woodworth \etal \cite{7841040} further propose work that reduces the number of terms put in the index by extracting only the most important terms from an uploaded document, and extends the query using Wikipedia and Synonym to capture the broader meaning of the query. The result is a space-efficient index which still achieves an accurate semantic search.

Table~\ref{fig:se-approach} provides a summary of categorization of different \se~approaches in terms of components mentioned in Figure~\ref{fig:ovrarch}. We point out references to the works in each \se~type, so readers could easily compare the difference between these works.

\begin{table*}[htbp]
\caption{\label{fig:se-approach}Categorizing different studied \se~systems in terms of components they contain.}
\centering{
\resizebox{\textwidth}{!} {
\begin{tabular}{|l | c | c | c | c | c | c }
  
\hline
\textbf{Search Approach} & \textbf{\parbox{2cm}{File \\ Extractor}} & \textbf{\parbox{2cm}{\small{Using \\ Index File}}} & \textbf{\parbox{2cm}{Multi-keyword\\ Searching}} & \textbf{\parbox{2cm}{Query Expansion\\ }} & \textbf{\parbox{2cm}{Ranking\\ Search Result}}\\
  \hline
  \hline
\textbf{Keyword Search} &   N/A    &  \cite{goh2003secure,liu2015efficient,wang2012enabling,wang2010secure,homomorphic:indexing:ren}     & \cite{5935306,6413690}     & N/A   & \cite{wang2012enabling,wang2010secure} \\ \hline
\textbf{Regular Expression Search} & N/A   & N/A   & N/A  & \cite{amini14,aminispe17}   & \cite{moataz2013semantic,song2000practical,wang2010secure,chang2005privacy,goh2003secure,boneh2007public}  \\ \hline
\textbf{Semantic Search} & \cite{7841040, fu2016toward,sun2014privacy,saleem2017secure,wang2012enabling}        & \cite{7841040,sun2014privacy,saleem2017secure,wang2012enabling}   &   \cite{7841040}     & \cite{7841040, fu2016toward,li2010fuzzy}   & \cite{7841040,sun2014privacy,xia2014secure,moh2014efficient}  \\ 
\hline
\hline

\end{tabular}
}}
\end{table*}

\section{Security Analysis of Searchable Encryption Systems}\label{sec:shortcoming}
Based on security criteria of \se~systems (mentioned in Section~\ref{subsec:crit}) and the level of information leakage, we can categorize current \se~systems into four security levels. 
The differences between these security levels are summarized in Table \ref{fig:sec-level} and are explained below.

\begin{table*}[htbp]
\caption{\label{fig:sec-level}Security level of current secure search systems.}
\centering{
\resizebox{\textwidth}{!} {
\begin{tabular}{l | c | c | c | c | c | c | c}
  
\hline
\textbf{Security Level} & \textbf{\parbox{2cm}{Leak Plain\\ Text Document}} & \textbf{\parbox{2cm}{\small{Using Trusted Computational Base}}} & \textbf{\parbox{2cm}{Leak Index\\ Data}} & \textbf{\parbox{2cm}{Plain\\ Data Stored Remotely}} & \textbf{\parbox{2cm}{Leak Access\\ Pattern}} & \textbf{\parbox{2cm}{Leak Search\\ Pattern}} & \textbf{\parbox{2cm}{Related\\ Works}}\\
  \hline
  \hline
\textbf{Somewhat secure}   & No       & Yes    & No   & Yes    & Yes     & Yes   & \cite{sun2014privacy,amini14,wang2013efficient} \\ \hline
\textbf{Semi secure} & No          & No    & Partially   & No    & Yes     & Yes   & \cite{7841040,saleem2017secure} \\ \hline
\textbf{Secure} & No          & No    & No   & No    & Yes     & Yes   & \cite{moataz2013semantic,song2000practical,wang2010secure,chang2005privacy,goh2003secure,boneh2007public}  \\ \hline
\textbf{Fully secure} & No          & No    & No   & No    & No     & No   & \cite{goldreich1996software}  \\ 
\hline
\hline

\end{tabular}
}}
\end{table*}

\paragraph*{Somewhat Secure} The \se~systems in this category often deploy a trusted server (also known as a private cloud~\cite{sun2014privacy} or a gateway~\cite{amini14}) in between the third-party server (\eg public cloud) and the client device. The \se~systems that leverage edge/fog computing~\cite{yi2015security} paradigms also fall under this category.

In Somewhat Secure \se~systems, unencrypted documents are sent to the private cloud, where they are parsed and encrypted to form the index structure. These systems not only leak access and search patterns, but  also expose the documents' contents to an internal attacker  of the trusted server. Sun \etal \cite{sun2014privacy} make use of the computational capacity of private clouds to build a Semantic Relations Library (SRL). The SRL quantifies the semantic distance of each term to other terms based on their co-occurrence in the dataset. However, SRL is maintained unencrypted on the private cloud, hence, is susceptible to attacks on the private cloud.

\paragraph*{Semi Secure} In \se~systems with this level of security, the auxiliary index structure is partially encrypted. That is, some information about the documents or the keywords are not encrypted and can be leaked from the index structure.

In \cite{saleem2017secure}, the index structure keeps track of encrypted keywords and for each keyword, the list of documents that include the keyword. However, the weight (\ie score) of the keyword in each document is maintained in an unencrypted manner. Therefore, the index structure is partially encrypted and reveals the relevance score that represents the relatedness of a given keyword to a document. In \cite{7841040}, Woodworth \etal provided a semantic search system for encrypted data in the cloud. Their system, named S3C, also exposes the frequency of a keyword within the original documents. These unencrypted data are considered security holes in the \se~system that attackers can take advantage of and conduct a statistical attack.

\paragraph*{Secure} Such \se~systems do not trust any part of the system, except the client's device. Also, the auxiliary index is properly secured and does not expose any plain text data to the server. Keywords in the index structure are hashed using SHA-1 or HMAC-SHA~\cite{wang2010secure,hahn2014searchable} methods. Also, relevance scores are encrypted using OM-OPSE method (as explained in Section~\ref{omopse}). 

Because of the aforementioned reasons, Secure \se~systems are less prone to internal attacks, in compare to those in the Somewhat or Semi Secure categories. However, the Secure \se~systems still leak the access and search patterns. Although there are methods (\eg Private Information Retrieval \cite{kushilevitz1997replication,boneh2004public}, Oblivious RAM \cite{goldreich1996software}) to avoid such leaks, they slow down the search process. Thus, Secure \se~systems prefer to expose these leaks in favor of the search performance. Instances of Secure \se~systems can be found in \cite{moataz2013semantic,song2000practical,wang2010secure,chang2005privacy}. 

\paragraph*{Fully Secure} The \se~systems in this category provide full security of the data and the search operation. Systems in this category do not even reveal the search and access pattern to the server.

Research undertaken by Goldreich \etal \cite{goldreich1996software} falls into the fully secure category. It provides search and access pattern security through ORAM method (see Section~\ref{sec-shortcoming}). This strong security, however, comes with poly-logarithmic rounds of communication between the server and the user, which restrains its usability. In the same research, the authors introduced another method that needs only two-round communication between the users and the server. Nonetheless, it induces a square root complexity overhead in the set up phase. 

\section{Application Driven Domain}\label{sec:app}
Certain regulations \cite{king2012protecting} require some businesses and industries (\eg health care) to store their data in a secure a manner. As such, they need to store their data in an encrypted format, if they were to use a third-party servers or a public cloud for storage. Therefore, there have been several efforts from those businesses to tailor \se~solutions for their specific domains.  This section provides an overview of such works. In the rest of this section, we investigate such domain-specific \se~systems.

\paragraph*{Health care} For health care providers, all Personal Health Records (PHRs) have to be encrypted to guarantee a patient’s privacy \cite{li2013scalable}. Health care providers focus on maintaining the privacy of PHR in an emerging patient-centric model that uses public clouds to easily store, retrieve, and share health information between medical centers. Although the public cloud solutions are convenient and attractive for health care providers, privacy concerns restrain its potential \cite{yang2015hybrid}. 

The owner of the PHR has the utmost right to share and distribute the decryption key to other users for personal or professional purposes \cite{li2013scalable}. However, the encrypted PHR makes key functionalities, such as keyword search by multiple users challenging. Multiple data users need to request for the decryption key to access the encrypted PHR. However, the key requests by users are generally unpredictable and this makes it challenging for the data owner to manage key distribution, as they are not always online \cite{li2013scalable}. To overcome this challenge, attribute-based encryption (ABE) scheme is proposed~\cite{goyal2006attribute}. In ABE, there is a set of attributes that manage the access policies. Only users with the valid decryption key that matches the attribute can decrypt the patient's PHR and users with a revoked key cannot do anything with the encrypted PHRs. The ABE encryption scheme also enables a patient to exchange PHRs with a group of users without prior knowledge of the complete list of users. 

In \cite{ali2017privacy}, Naseeruddin \etal developed an ABE-based schema that allows the system to audit user access, allowing it to identify the source of a breach and detect if information is distributed properly or if there was illegal access to the data. An encrypted PHR will be delegated to $n$ trusted authorities from the data owner (patient). To access and search on the encrypted data, a data user need to acquire the approval of $k$ trusted authorities. Their work modified ABE with a signature threshold $A(k,n)$ where a valid message is guaranteed if there are at least $k$ valid signature shares out of $n$ verified parties. In the Key generation algorithm of ABE, the key is generated based on an ABE-modification to encrypt the data before outsourcing it to the public cloud, which can be searched later to retrieve necessary information. 

Another modification of ABE is proposed in \cite{li2013scalable,li2011authorized} to add another layer of fine-grained privacy protection beyond the underlying cryptographic mechanisms to enable \se~or data access control. The study develops a hierarchical predicate encryption (HPE), in which the data owner generates a key and distributes it to a local trusted authority (LTA). These LTAs are able to delegate the key to allow authorized users to search to that patient's PHR. In the search phase, the server has to verify that the user has a valid signature from a registered LTA before performing search. It is worth noting that this study only supports keyword search, and not other forms of search.%[Done] [Mohsen: revise this partagraph]

These methods are appropriate for normal medical procedures, but may hamper first-aid treatment in emergency situations in which a patient's life is at risk and an authorized user is not present. To combat this challenge, Yang \etal \cite{healthcare:yang} implement adaptive dual-layer access control, in which normal approved users are allowed to see all of the patient's health care data, while outside users can view a subset of the data in emergency situations using a password-based break-glass access mechanism. This approach provides data security while limiting danger in emergency scenarios.

\paragraph*{Law Enforcement} Other industries, such as law enforcement, %[Done][<-- Mohsen: orrganizations??? do you mean law enforcement?], 
can potentially have privacy restrictions in granting access to their datasets. Data owners, in this industry, require data that is stored in external sources (\eg in a cloud) to be encrypted, and \se~systems may only be used to enable search over data without providing direct access to the data. For example, a detective needs to know if the subject has any matching record in the court reports. The detective but does not need to (or cannot) access the information of all people in the court system.

\paragraph*{Text Editors} Li \etal \cite{li2010fuzzy} propose a secure search system which could handle user typos through a fuzzy keyword search. Wang \etal \cite{wang2013efficient} used a similar approach to find matches for similar keywords to the user's query by using edit distance as a similarity metric, allowing for words with similar structures and minor spelling differences to be matched. However, these methods reveal the topic of the external sources that an attacker can use to categorize the data sources based query retrieval. 

\paragraph*{Other Domains} Financial and military domains generally have strict data privacy and confidentiality rules \cite{anton2004financial}, however,  currently, there is no dedicated study on the particular use of \se~in these domains.

\section{Future Research Directions}
\paragraph*{1. Clustering Encrypted Data}
%Since the idea of searching encrypted data introduced in late 90's, many \se~systems have been developed. Recent works in this area primarily try to improve the search accuracy, performance, and security. Despite provable improvements in security, \se~solutions still need to enhance their performance and maintain the real-timeness of the search operation. This is particularly important, as datasets are dramatically growing in size and more businesses are looking to outsource their big data to the cloud. One potential and promising approach to deal with big data \se~is to cluster big data and navigate the search operation only to clusters relevant to the search query. As the semantics of keywords are lost for encrypted data, clustering encrypted data is challenging. Once the clustering is done on encrypted data, the remaining challenge is how to proactively navigate the search operation only to relevant clusters while the data are encrypted? Efficient number of clusters for a given dataset and the ways to keep even clusters are other challenges in clustering of encrypted data. Little research have been undertaken on clustering of the encrypted data. Therefore, this area is one of the demanded research avenues for future \se~systems and it needs further exploration both from academia and industry.
Since the field's inception, \se~systems have primarily strived to improve search accuracy, performance, and security.  Despite many provable improvements to security, \se~solutions still need to enhance their performance and maintain the real-timeness of the search operation. This is particularly important as datasets get substantially larger, and with more businesses looking to outsource their big data to the cloud. One promising approach to dealing with big data is to cluster the data and restrict the search to relevant clusters.  This approach presents several challenges: clustering encrypted keywords based on some meaningful semantic data, determining which clusters are most relevant at seasrch time, and determining how many of those clusters should be searched. 

\paragraph*{2. Searchable Encryption Across Multi-source Datasets}
As datasets grow larger and more organizations opt to use remote cloud storage, it becomes increasingly likely that users and organizations will need to search across multiple sources. For example, a law enforcement officer who needs to obtain information on a suspect may need to search across datasets from other jurisdictions.
%One direction that \se~systems are growing and need further research exploration is to deal with multi-source datasets. In fact, many \se~systems, nowadays, need to perform search over data that belong to different organizations. For instance, consider a law-enforcement officer who needs to obtain criminal background of a subject. The officer needs a \se~system to enable transparent search across multiple datasets (\eg from jail reports, court reports, social networks, and department of motor vehicle). 
One challenge to this is that each of these organizations can potentially have their own privacy and confidentiality policies and use various encryption standards. The datasets may also have different characteristics, such as document structure or length. Further research is needed to deal with a wider variety of data, and combining multiple levels of security.
%One challenge to this is that each of these organizations can potentially have their own privacy and confidentiality policies and use various encryption standards. Further research is required to overcome these discrepancies on the datasets and enable a transparent search system across all datasets in a real-time manner. Another challenge is that these datasets have different characteristics. For instance, the court dataset may include reports with several pages whereas the social network dataset includes documents with only few keywords. This type of discrepancy requires various ways of dealing with different datasets, \eg different ways to extract keywords or to expand search queries. 

\paragraph*{3. Edge Computing for Searchable Encryption}
Most current \se~systems have a two-tier architecture, utilizing primarily a user-end and cloud-end. The user-end includes the client machine, often in the form of a thin-client (\eg smart phone or tablet) with limited computational power. This often creates a performance bottleneck, as much of the document and query pre-processing must occur on the client-end since it is the only trusted component.
%Most of the current searchable encryption systems, have a two-tier architecture, namely user-end and server-end (also known as cloud-end). The user-end includes the client machine which is usually in form of a thin-client (\eg smart-phone or a tablet) with a limited computational and power capabilities. However, as the client machine is generally the only trusted computational base in the \se~system, it takes the burden of pre-processing data (\eg encrypting documents and expanding search queries). However, these pre-processing tasks make the client machine the bottleneck of the \se~system. 
An edge computing paradigm can be utilized to bypass this bottleneck. Use of this paradigm would add additional architectural tiers along the network between the server and client. Off-loading computational work to an edge node can substantially improve performance, but weakens security, as edge nodes are more vulnerable to attacks than clients. Further research efforts should determine what processes can be performed on edge nodes while minimally compromising security.
%Edge computing paradigm can be deployed for \se~systems to bypass the bottleneck. In this paradigm, in addition to the public cloud-tier, another computational tier is considered at the edge of the network, near to the users and sources of data. Security-wise, in edge computing, the farther we get from the client machine, the less trustworthy it becomes. That is, the trustworthiness of edge tier is between the client tier (the fully trusted one) and the public cloud (the least trusted one). By efficiently deploying \se~systems on the edge computing paradigm, some of the client-side processing tasks can be carried out on the edge nodes. Further research efforts needs to determine the processes that can be carried out on the edge tier to speed up the \se~systems. In addition, changes that needs to be applied on the client processes to adopt the less secure nature of edge-tiers have to be investigated.

%\paragraph*{4. Domain-Specific Searchable Encryption}
%Searchable encryption systems in different domains deal with diverse data types with distinct security constraints. For example, a financial institution needs a search mechanism for their encrypted data that also ensures the transparency for audits. Another example is a military system where users' ranks imply different access levels to search and access encrypted documents. These problems bring up challenges that can be specific to that domain. Solving these domain-specific challenges helps the different industries to embrace cloud services and unleash their potentials for \se. 

\paragraph*{4. Utilizing Blockchain in Searchable Encryption}
With recent breakthroughs, blockchain technology and decentralization idea nowadays are becoming more promising aspects and gaining more attention from the \se~community. There are several of proposals to utilizing blockchain in \se. Hu \etal \cite{hu2018searching} introduce decentralized privacy-preserving search scheme leveraging smart contract where blockchain verifies the accuracy and fairness in the contract between users and cloud provider. In other research, Cai \etal \cite{7996810}suggest decentralized storage platform which is an concept of people leasing their hardware capacity as a service. However, it is questionable that outsourcing data to other storage providers could be exposed threats of data integrity which is mentioned in \cite{basu2018cloud} and thus affect autonomous payment of people involving the service \cite{7996810}. The system proposed by Cai \etal \cite{7996810} utilizes cryptocurrencies and cryptographic incremental hashing to solve the mentioned concerns. This is one of the very first works integrating blockchain with \se~systems which starts a promising direction. Nevertheless, it would need more investigation and experiments to develop more mature approaches in the future.

\section{Summary}\label{sec:conclsn}
In this work, we surveyed current \se~systems and techniques that perform various forms of search operation on encrypted data located on an untrusted third-party cloud. We identified common components of the \se~systems and the overall architecture through which these components interact with each other. Then, we provided a taxonomy that categorizes the \se~systems based on the type of search operation supported by them. We analyzed the security of the current \se~systems and categorized them into four security levels. We investigated particular demands of \se~systems in various domains, such as health care and law-enforcement. Finally, we offered suggested directions for future research, including clustering data for improved search speed, searching across multiple datasets, and utilizing edge computing and blockchains.

\vspace{-12pt}
\section*{Acknowledgments}
We would like to acknowledge anonymous reviewers of the manuscript. 
This research was supported by the Louisiana Board of Regents under grant number LEQSF(2017-20)-RD-B-06, and Perceptive Intelligence, LLC.

\vspace{-12pt}
\section*{References}

\bibliographystyle{elsarticle-num}
\bibliography{PreemptingPolicy}

\end{document}